\documentclass[11pt]{article}
\usepackage[numbers,round]{natbib}
\usepackage{hyperref}
\usepackage{amsmath,amssymb,euscript,yfonts
,psfrag,latexsym,dsfont,graphicx,bbm,color,amstext,wasysym,balance,mathtools}
\usepackage{subfig,float}
\usepackage{multirow}
\usepackage{bm}
\usepackage{authblk}
\usepackage{setspace}

\usepackage[margin=1in]{geometry}

\providecommand{\keywords}[1]{\textbf{Key words:} #1}
\newtagform{brackets}{[}{]}
\usetagform{brackets}
\makeatletter 
\renewcommand\@biblabel[1]{#1.} 
\makeatother

\begin{document}

\title{High-fidelity, accelerated whole-brain submillimeter in-vivo diffusion MRI using gSlider-Spherical Ridgelets (gSlider-SR)}

\author[1]{Gabriel Ramos-Llord{\'e}n \thanks{\textbf{Corresponding author:}\\
		Gabriel Ramos-Llord{\'e}n\\
		Psychiatry Neuroimaging Laboratory,\\
		Department of Psychiatry, \\
		Brigham and Women's Hospital, Harvard Medical School,\\
		02215 Boston, Massachusetts, USA,\\
		Telephone: +1 617-525-6124, \\
		Email: gramosllorden@bwh.harvard.edu}}
\author[1]{Lipeng Ning}
\author[2]{Congyu Liao}
\author[1,3]{Rinat Mukhometzianov}
\author[3]{Oleg Michailovich}
\author[2]{Kawin Setsompop}
\author[1]{Yogesh Rathi}
\affil[1]{Department of Psychiatry, Brigham and Women's Hospital, Harvard Medical School, Boston, Massachusetts, USA}
\affil[2]{Athinoula A. Martinos Center for Biomedical Imaging, Massachusetts General Hospital, Harvard Medical School, Boston, Massachusetts, USA}
\affil[3]{Department of Electrical and Computer Engineering, University of Waterloo, Waterloo, Ontario, Canada} 
\date{}

\maketitle

\noindent
\\[.3cm]
{\noindent \textbf{Word count:} 4840 }\\[.4cm]
	
{{\Large{This manuscript is an old version. \\ Latest version, which is published in the journal Magnetic Resonance in Medicine, can be found at  \\ \url{https://doi.org/10.1002/mrm.28232} }	
\newpage
\begin{abstract}
	\noindent
	{\bf Purpose:} To develop an accelerated, robust, and accurate diffusion MRI acquisition and reconstruction technique for submillimeter whole human brain in-vivo scan on a clinical scanner. \\
	{\bf Methods:} We extend the ultra-high resolution diffusion MRI acquisition technique, gSlider, by allowing under-sampling in q-space and Radio-Frequency (RF)-encoded data, thereby accelerating the total acquisition time of conventional gSlider. The novel method, termed gSlider-SR, compensates for the lack of acquired information by exploiting redundancy in the dMRI data using a basis of Spherical Ridgelets (SR), while simultaneously enhancing the signal-to-noise ratio.  Using Monte-Carlo simulation with realistic noise levels and several acquisitions of in-vivo human brain dMRI data (acquired on a Siemens Prisma 3T scanner), we demonstrate the efficacy of our method using several quantitative metrics.   \\
	{\bf Results:} For high-resolution dMRI data with realistic noise levels (synthetically added), we show that gSlider-SR can reconstruct high-quality dMRI data at different acceleration factors preserving both signal and angular information. With in-vivo data, we demonstrate that gSlider-SR can accurately reconstruct 860 ${\mu}m$ diffusion MRI data (64 diffusion directions at b = 2000 $s/ {mm}^2$), at comparable quality as that obtained with conventional gSlider with four averages, thereby providing an eight-fold reduction in scan time (from 1 h 20 min to 10 min). \\
	{\bf Conclusion:} gSlider-SR enables whole-brain high angular resolution dMRI at a submillimeter spatial resolution with a dramatically reduced acquisition time, making it feasible to use the proposed scheme on existing clinical scanners.
\end{abstract}

\section{Introduction}
Diffusion MRI (dMRI) is a non-invasive imaging modality that permits the characterization of tissue microstructure as well as the structural connectivity of the human brain \cite{Basser1994,Assaf2008}. As it is sensitive to neural architecture, it is increasingly being used in the clinical investigation of several brain disorders \cite{Shenton2012, Thomason2011}. dMRI holds the promise of being a key tool to explore and understand the human brain at an unprecedented level of detail. In the quest for a rich and detailed understanding of the human brain, it is the image resolution of dMRI that is the main limitation and is the focus of active research of the MRI community.  Indeed, in clinical settings, diffusion-weighted images (DWI) are typically acquired at an isotropic resolution of 2 $mm$ \cite{Setsompop2018,Ning2016}. This causes undesirable partial volume effects, especially at the interface of different tissue types, such as the gray and white matter boundary. The limited resolution also has a significant impact on both white and gray matter studies. For example, in the superficial white matter regions, there is an abundance of short cortical association fibers (U-fibers) that connect cortical regions between adjacent gyri, which are difficult to trace at current resolutions. Increasing the current dMRI resolution will reduce the large partial volume effects, and facilitate the analysis of small structures that remain ``hidden"  at isotropic resolution of 2 $mm$ \cite{Ning2016}.

Unfortunately, increasing the dMRI resolution, or equivalently, reducing the voxel size, is challenging since the Signal-to-Noise-Ratio (SNR) is proportional to the size of the voxel \cite{Edelstein1986}. A 1 ${mm}$ isotropic dMRI acquisition will have eight times lower SNR compared to a 2 $mm$ isotropic acquisition. While the SNR can be increased using multiple acquisitions and averaging, nevertheless, because the SNR is proportional to the square root of the number of averages, at least 64 repetitions would be needed to match the SNR of a 2 ${mm}$ isotropic acquisition \cite{Ning2016}. Naturally, the total acquisition time of this averaging protocol is prohibitive, making this kind of approach impractical for human in-vivo settings.

The trade-off between image resolution, SNR, and acquisition time can be circumvented using super-resolution based methods.  Super-resolution methods fall in the category of reconstruction frameworks where a high-resolution image is estimated from a set of low-resolution images, each one sampled with different geometric schemes. Formulated as the solution of an inverse problem, the restored image (or the super-resolution image), suffers less from SNR penalty than a direct isotropic acquisition since the SNR of the low-resolution images, normally, thick-slices images, is substantially higher. As the total acquisition time of the set of thick-slice images is comparable to that of the direct isotropic acquisition, super-resolution based methods effectively break the trade-off between image-resolution and SNR. Super-resolution was first applied in the context of MRI in \cite{Peled2001}, and ever since has been used in a multitude of cases, e.g, anatomical MRI \cite{Greenspan2001,Greenspan2009, Gholipour2010,Poot2010},  quantitative relaxometry \cite{VanSteenkiste2017}, and diffusion MRI \cite{Peled2001,Scherrer2012,Poot2013,VanSteenkiste2016a,Ning2016}.  All of these super-resolution methods have been conceived and tested for resolutions $>$ 1 ${mm}$. Recently, a multi-shot and multi-slab acquisition sequence was proposed in \cite{chen2013robust, bruce20173d} for submillimeter dMRI acquisition. This scheme has the advantage of good SNR, but at the cost of long acquisition times (about 30 min of scan time for only 12 gradient directions) along with complications from inter-slab registration due to head motion. 

A turning-point in submillimeter diffusion MRI acquisition was the introduction of Slice Dithered Enhanced Resolution (Slider) \cite{vu2018evaluation, Ning2016} and recently the Generalized Slice Dithered Enhanced Resolution (gSlider) method \cite{Setsompop2018}.  gSlider is an acquisition framework that utilizes a novel radio frequency (RF) encoding basis to excite multiple slabs of the whole human brain simultaneously, and then combine them to create a high-resolution thin-slice reconstructed volume.  Specifically,  several RF slab-encoded volumes are acquired in consecutive TRs,  each volume encoded by a given ``component"  of what is called an RF-encoding basis, i.e., encoded by a particular RF excitation profile selected from a group of predefined RF waveforms (the basis). The elements of the RF-encoding basis are specifically designed to be highly independent between each other along with having a larger slice thickness to allow for increased SNR.  When the simultaneously acquired slabs are un-aliased with blipped-CAIPI, the set of RF-encoded thick-slice diffusion-weighted imaging (DWI) volumes are used to reconstruct a super-resolution thin-slice DWI volume using standard Tikhonov regularization \cite{Setsompop2018}. gSlider has been successfully applied to reconstruct diffusion MRI data with spatial isotropic resolution ranging from 660 ${\mu}m$ to 860 ${\mu}m$ with b-values between 1500 $s/mm^2$ and 1800 $s/mm^2$ \cite{Setsompop2018,Wang2018,Liao2019}.

To obtain high SNR and high angular resolution dMRI data, the current gSlider protocol requires long acquisition time. However, the redundancy within the RF-encoding as well as the different gradient directions (or q-space points) can be exploited to reduce the acquisition time dramatically to make it clinically feasible for whole-brain (submillimeter) in-vivo acquisition. In this work, we exploit this redundancy and propose an algorithm (termed gSlider-SR) to reconstruct high SNR gSlider data using a basis of spherical ridgelets (SR), which has been shown to be a highly sparse basis for dMRI data reconstruction, thereby allowing large undersampling factors \cite{Michailovich2009, Michailovich2010, Michailovich2011a, Rathi2014, Michailovich2014, ning2015sparse, NING2016386}. Specifically, for each diffusion direction, we propose to excite the slab-encoded volumes with just a subset of the RF-waveforms that constitute the RF-encoding basis of gSlider. Therefore, contrary to conventional gSlider, where each slab-encoded DWI volume that is probed at a given q-space point is encoded with all of the components of the RF-encoding basis, here we only employ a small subset of that basis.  To provide complementary diffusion information, the subset of RF-encoding components is not static but varies along different q-space points. Reconstructing thin-slice DWI volumes from such an under-sampled set of measurements is an ill-posed, super-resolution reconstruction problem. However, we can cast this problem as that of reconstructing the entire dataset from a sparse set of measurements. Equipped with spherical ridgelets (SR), we recover the thin-slice DWI set by solving a constrained $l_1$ minimization problem whose theoretical background relies on the theory of Compressed-Sensing. We validate gSlider-SR both qualitatively and quantitative with Monte-Carlo (MC) based simulations and in-vivo human brain data, where we showcase an accurate reconstruction of 64  DWI volumes (b = 2000 $s/mm^2$) with 860 ${{\mu}m}$ isotropic resolution in a scan time of approximately ten minutes.

An earlier version of the proposed gSlider-SR has been presented as an abstract at ISMRM 2016 \cite{Ning2016ISMRM}.  The optimization algorithm presented in this work is largely based on the reconstruction framework introduced in \cite{Ning2016,Ning2016ISMRM}. However, there are several important differences and new contributions in this paper. First, the work in \cite{Ning2016} used the Slider basis, which is substantially different than the optimized gSlider basis used in this work. In particular, the Slider basis consisted of multiple overlapping thick-slice dMRI volumes, whereas in our approach, the set of thick-slices are acquired in a more complex setting, each one encoded with a novel slab RF-encoding technique \cite{Setsompop2018}.  The RF-encoding basis scheme offers more versatility and robustness to motion than the technique based on overlapping thick-slices. Consequently, in this work, we adapt the reconstruction framework for this new gSlider basis. Second, the work in \cite{Ning2016} was an initial proof of concept, whereas in this work, we perform a comprehensive set of simulations to test the performance of gSlider-SR with various noise levels and at different under-sampling factors. We also extend the validation of the method on real in-vivo data using several quantitative metrics and compare it comprehensively with 4-averages of the gSlider acquisition, thereby comparing whole-brain data quality.  Thus, this work significantly enhances and builds on the initial framework developed in \cite{Ning2016}.

\section{Theory}
\subsection{Conventional gSlider reconstruction} 

In this section, the basic theory of conventional gSlider based acquisition and reconstruction is covered. We also introduce the mathematical notation that will be used throughout the rest of the paper. 
In what follows, it is assumed that the simultaneously acquired gSlider slabs have been unaliased with parallel imaging and transformed into image space with any coil-combine technique available at hand. The contiguous concatenation of the unaliased slabs along the slice-direction constitutes a thick-slice volume. The set of thick-slice volumes acquired at given q-space points $\bm q_j$ ($j=1, ..., N_q$) is named here as a thick-slice DWI set. The thick-slice DWI set that is encoded with the $k$-th component of the RF-encoding basis, i.e., excited with the $k$-th RF-waveform is denoted as $\bm Y_k \in \mathbb{R}^{N_{LR} \times N_q}$.  We direct the reader to \cite{Setsompop2018} for more details about the RF-encoding basis. The number $N_{LR}$ stands for the total number of voxel in each DWI volume and is given by $N_{LR} = n_x \times n_y \times n_\text{z-slices}$, where $n_x \times n_y$ is the size of the in-plane matrix, and $n_\text{z-slices}$ is the number of thick slices that are acquired along slice-encoding direction. Such thick-slice DWI sets, $\bm Y_k$ are related by means of a forward-model to an unobserved isotropic, high-resolution thin-slice DWI set that we denote  as $\bm S \in \mathbb{R}^{N \times N_q}$, where the total number of voxels, $N$, is defined as $N = n_x \times n_y \times n_z$. The number of thin slices along direction $z$, $n_z$, is given by  $n_z = \text{AF} n_\text{z-slices}$, where AF $\geq 1 $ the ratio of the slice thickness of $\bm Y_k$ over that of $\bm S$.
Finally, the forward-model that connects $\bm S$ to $\bm Y_k$ can be  written as:
\begin{equation}
\bm Y_k = \bm D_k \bm S + \bm \eta_k \quad \text{with } k=1,...,N_{RF},
\label{eq:linearfully}
\end{equation}
where  $\bm D_k \in \mathbb{R}^{N_{LR} \times N}$ is the gSlider downsampling operator that corresponds to the $k$-th RF-encoding basis (the number of components of the basis is $N_{RF}$) and where the term  $\bm \eta_k$ represents random noise of the acquisition.  Reconstructing $\bm S$ from the low-resolution data $\bm Y_k$ is an inverse problem, in particular, a superresolution reconstruction problem. In conventional gSlider,  $N_{RF} = \text{AF}$, which implies that the number of unknowns is the same as the number of measurements. In this scenario, a simple Tikhonov regularization produces satisfactory results \cite{Setsompop2018}.  As mentioned in the introduction, while conventional gSlider is SNR-efficient compared to a full isotropic high-resolution acquisition, it, however, requires a  relatively long-acquisition time. One way to shorten the acquisition time is by reducing the number of components of the RF-encoding basis.  The next section is devoted to further elaborate on this undersampled gSlider RF-encoding data scenario.  

\subsection{Undersampled gSlider RF-encoding data} 

In this work, we shorten the acquisition time of gSlider  by encoding each of the DWI of  $\bm S$ with a subset of the $N_{RF}=5$ RF-encoding basis. To illustrate this scenario, let us focus on Fig. \ref{fig:scheme}. In conventional gSlider, all the diffusion-weighted volumes, say $N_q = 64$ in total, are excited with all the five RF-waveforms (see Fig.\ref{fig:scheme}.(a)), i.e., a total of 320 volumes are acquired. Within an undersampled RF-encoding data scenario ( Fig.\ref{fig:scheme}.(b)), an incomplete RF-encoding basis is used to excite  DWI volumes probed at different q-space points. DWI volumes synthesized with ``red" q-space points (32 in total)  are encoded with just the first, third and fifth RF-encoding component. On the other hand, the fourth and fifth RF-encodes are used to excite only DWI volumes that are acquired with ``blue" q-space points, i.e., the remaining 32 diffusion directions. Note that the complete set of $N_q = 64$ diffusion directions are encoded at least once, but for each RF-encoding profile, only $N_{q_k}=$ 32 DWI volumes are excited. This represents a 2 X acceleration (as a total of only 160 DWI volumes are acquired) compared to a standard gSlider acquisition (with 320 DWI volumes).

***FIG. 1 APPEARS NEAR HERE***\\

Mathematically, this scenario can be described as follows:
\begin{equation}
\bm Y_k = \bm D_k \bm S \bm \Omega_k + \bm \eta_k \quad \text{with } k=1,...,N_{RF},
\label{eq:linear_under}
\end{equation}
where $\bm \Omega_k$ ($N_q \times N_{q_k}$)  is a sampling  (binary) mask which determines whether  a given diffusion direction has been encoded by the $k$-th RF profile. Recovering $\bm S$ from $\bm Y_k,  k=1,...,N_{RF}$,  in Eq.\ref{eq:linear_under} is an ill-posed, super-resolution reconstruction problem, where many solutions exist. Solving this undetermined reconstruction problem demands more advanced techniques than a simple Tikhonov regularization. It, however, becomes tractable if prior knowledge about the structure of $\bm S$ is incorporated into the reconstruction framework.  In this work, such a prior knowledge comes in form of spherical ridgelet (SR) basis \cite{Michailovich2009}, that we present briefly below.

\subsection{Spherical ridgelets basis for diffusion signal recovery} 

Spherical ridgelets (SR) were proposed originally in \cite{Michailovich2009}, and have been successfully applied for diffusion signal recovery in many scenarios  \cite{Michailovich2009, Michailovich2010, Michailovich2011a, Rathi2014, Michailovich2014, ning2015sparse, NING2016386}.  In a nutshell,  spherical ridgelets are functions defined on the unit sphere, $\Psi (\bm{q})$ ($\bm q \in \mathbb{S}^2$), that are designed to represent any dMRI signal. Given a collection of spherical ridgelets, ${\{\Psi_m\}}_{m=1}^M$, any function $s(\bm q)$ with $\bm q \in \mathbb{S}^2$  can be written as 
\begin{equation}
s(\bm q) = \sum_{m=1}^M c_m\Psi_m(\bm q), 
\end{equation}
where $c_m$, $m=1,...,M$, are the SR coefficients.  Let the column vector $\bm s_n \in \mathbb{R}^{N_q}$ denote the diffusion signal observed at voxel $n$ of the  high-resolution DWI set $\bm S$, i.e., $\bm S =  [\bm s_1, \bm s_2,  \cdots,  \bm s_N ]^T$. 
It is then possible to write $\bm s_n = \bm A\bm c_n$, with
\begin{equation}
\bm A = \begin{pmatrix} \Psi_{11} & \Psi_{12} & \cdots & \Psi_{1M} \\
\Psi_{21} & \Psi_{22} & \cdots & \Psi_{2M} \\
\vdots & \vdots &\vdots & \vdots \\
\Psi_{N_q1} & \Psi_{N_q2} & \cdots  & \Psi_{N_qM} 
\end{pmatrix},
\end{equation}
and  where $\Psi_{jm}$ is the $m$-th spherical-ridgelet function evaluated at q-space point $\bm q_j$, and $\bm c_n \in \mathbb{R}^M$ is the vector of SR coefficients at voxel $n$. It should be noted that the spherical ridgelets form an over-complete basis. As noted earlier, the dMRI signal can be represented in a sparse manner in the SR domain, thereby satisfying the theoretical guarantees for robust signal recovery from sparse measurements \cite{candes2011compressed}.
Compressed-sensing theory asserts that, even with a very low number of measurements,  accurate estimation of $\bm c_n$, and hence $ \bm S$,  is possible if $\bm c_n$ is sparse and if the spherical ridgelets representation matrix $\bm A$ is incoherent with respect to the diffusion sampling operator $\bm \Omega_k$.  Robust signal recovery is obtained by solving an $l_1$ norm-based minimization approach to estimate $\bm c_n$ as described in the next section.

\section{Methods}
\subsection{gSlider-SR:  accelerated gSlider-Spherical Ridgelets reconstruction} %%4h
\subsubsection{gSlider-SR as a constrained, $l_1$-minimization problem}
An estimate of the superresolution DWI set  $\bm S$ of Eq.\ref{eq:linear_under}, $\hat{\bm S}$, is obtained as the solution of the following constrained $l_1$-minimization problem:
\begin{align}
& \min_{\bm S, {\{\bm c_n\}}_{n=1}^N}   \frac{1}{2}\sum_{k=1}^{N_{RF}} \vert \vert \bm Y_k -\bm  D_k \bm S \bm \Omega_k  \vert \vert_2^2 + \lambda \sum_{n=1}^N{\vert \vert \bm c_n\vert \vert}_1 \nonumber \\ 
& \text{subject to  } \bm S = {[\bm A\bm c_1, \bm A\bm c_2,  \cdots,  \bm A\bm c_{N}]}^T,
\label{eq:problem}
\end{align}
where the $l_1$ penalty parameter, $\lambda$, controls the influence of the sparsity regularization over the data-fidelity term. The constrained problem of Eq.\ref{eq:problem} can be efficiently solved using the Alternating Direction Method of Multipliers (ADMM) algorithm \cite{Boyd2011,Ning2016}. First, the so-called augmented Lagrangian function is constructed:
\begin{equation}
 \frac{1}{2}\sum_{k=1}^{N_{RF}} \vert \vert \bm Y_k -\bm  D_k \bm S \bm \Omega_k  \vert \vert_2^2 + \lambda \sum_{n=1}^N{\vert \vert \bm c_n\vert \vert}_1 + \frac{\rho_1}{2} \sum_{n=1}^N {\vert \vert \bm s_n + \bm {\Lambda}_n - \bm A \bm c_{n}  \vert \vert}_2^2, \label{eq:Lagrangian}
\end{equation}
where $\bm \Lambda_n$ is an auxiliary variable and $\rho_1$ a regularization parameter. 
Minimizing the Lagrangian (Eq.\ref{eq:Lagrangian}) for $\bm S$ and ${\{\bm c_n\}}_{n=1}^N$ is equivalent to minimizing the following two subproblems. Given an estimate of the SR coefficients $ {\{\bm c_n^{(t)}\}}_{n=1}^N$ at iteration $(t)$, $\bm S^{(t+1)}$ can be obtained as a closed-form solution of following linear least squares (LLS) problem:
\begin{equation}
\min_{\bm S} \frac{1}{2}\sum_{k=1}^{N_{RF}} \vert \vert \bm Y_k -\bm  D_k \bm S \bm \Omega_k  \vert \vert_2^2 +  \frac{\rho_1}{2} \sum_{n=1}^N {\vert \vert \bm s_n + \bm {\Lambda}_n^{(t)} - \bm A \bm c_{n}^{(t)}  \vert \vert}_2^2.
\label{eq:LLS}
\end{equation}
On the other hand, to get an estimate of the spherical ridgelets coefficients, $ {\{\bm c_n^{(t+1)}\}}_{n=1}^N$, we can use the traditional basis pursuit (BP) algorithm for $l_1$ minimization:
\begin{equation}
\min_{{\{ \bm c_n\}}_{n=1}^N}  \frac{\rho_1}{2} \sum_{n=1}^N {\vert \vert \bm s_n^{(t+1)} + \bm {\Lambda}_n^{(t)}  - \bm A \bm c_{n} \vert \vert}_2^2 + \lambda \sum_{n=1}^N{\vert \vert \bm c_n\vert \vert}_1 
\label{eq:BPD}
 \end{equation}
 Next step is to update the auxiliary variable as follows:
 \begin{equation}
  \bm \Lambda_n^{(t+1)} =   \bm \Lambda_n^{(t)} + (\bm s_n^{(t+1)} - \bm A \bm c_{n}^{(t+1)}).
  \label{eq:multiplier}
 \end{equation}
 These three steps are sequentially repeated until convergence.
 \subsubsection{Total Variation for spatial sparsity}
 To exploit the sparsity in the spatial DWI image, we can also use a 3D Total Variation (TV) operator, which helps reducing noise while preserving structural details. In this work, the TV semi-norm of $\bm S$,  ${\vert \vert \bm S \vert \vert}_{TV}$, as defined in \cite{Ning2016} is used and the associated penalty parameter is denoted as $\lambda_{TV}$. The application of ADMM to this extended constrained reconstruction problem is straightforward and is provided in \cite{Ning2016}. We just show the resulting subproblems that constitute the main core of the final super-resolution reconstruction algorithm.
The algorithm now incorporates a new TV subproblem:
 \begin{equation}
\min_{\bm Z} \frac{\rho_2}{2}  {\vert \vert \bm S^{(t+1)}+ \bm {\gamma}^{(t)} - \bm Z \vert \vert}_2^2 + \lambda_{TV} {\vert \vert \bm Z\vert \vert}_{TV},
\label{eq:TV}
 \end{equation}
 where $\rho_2$ and $\lambda_{TV}$ are regularization parameters and $\bm Z$ and $\gamma$ are auxiliary variables. The solution of Eq.\ref{eq:TV} provides a denoised estimate of $\bm S$, i.e., $\bm Z^{(t+1)}$. No modifications in Eq.\ref{eq:BPD} are necessary to account for the TV term. However, the LLS problem of Eq.\ref{eq:LLS} should be slightly modified: 
 \begin{equation}
 \min_{\bm S} \frac{1}{2}\sum_{k=1}^{N_{RF}} \vert \vert \bm Y_k -\bm  D_k \bm S \bm \Omega_k  \vert \vert_2^2 +  \frac{\rho_1}{2} \sum_{n=1}^N {\vert \vert \bm s_n + \bm {\Lambda}_n^{(t)} - \bm A \bm c_{n}^{(t)}  \vert \vert}_2^2 +\frac{\rho_2}{2}  {\vert \vert \bm S+ \bm {\gamma}^{(t)} - \bm Z^{(t)} \vert \vert}_2^2.
 \label{eq:LLSTV}
 \end{equation}
Lastly, $\bm \gamma$  is updated as follows:
 \begin{equation}
 \bm {\gamma}^{(t+1)} = \bm {\gamma}^{(t)} + ( \bm S^{(t)} - \bm Z^{(t+1)} ).
 \label{eq:multiplier2}
 \end{equation}
 The algorithm is initialized with a Tikhonov regularization-based solution and terminated when either the $l_2$ norm between consecutive iterations is below a given tolerance $\epsilon$ or the number of iteration exceeds a given maximum number, $N_{iter}$.

\subsection{Experimental validation}
The proposed super-resolution reconstruction framework  was validated with simulated and in-vivo human brain data, both quantitatively and qualitatively.

\subsubsection{Simulation experiments } 
For a variety of q-space undersampling scenarios, thick-slice DWI sets $\bm Y_k$, $k=1,...,N_{RF} = 5$, were simulated following the forward-model of Eq.\ref{eq:linear_under}, where $\bm \eta_k$ represents uncorrelated zero-mean Gaussian noise, and the ground-truth (GT) high resolution DWI set $\bm S$ was created as described below.  \paragraph{Ground-Truth (GT) creation:} Whole human brain gSlider-SMS data was collected from a healthy male volunteer with a Siemens 3T Prisma scanner.  Four scans of the full brain ($\text{FOV} = 220 \times 220 \times 163 \text{ mm}^3$) were  obtained using the following parameters. With a single-shot EPI sequence, 38 thick axial slices (slice thickness = 4.3 mm) were acquired with matrix size = 256 $\times$ 256 and 860 ${\mu m}$ in-plane isotropic resolution, ${N_\text{RF}}=\text{AF}=5$ RF-encodings, Multi Band (MB) = 2, phase-encoding with undersampling factor $R_{\text{in-plane}} = 3$, partial Fourier = 6/8, TR / TE = 3500 / 81 ms, $N_q = 64$ diffusion directions  (b = 2000 s/${mm}^2$)  and 8 b0 images (non-diffusion weighted images). The $N_q = 64$ diffusion directions were approximately equally distributed over the hemisphere, $(x,y,z) \in \mathbb{S}^2 \text{ with }  y>0$.
The total acquisition time was about 1 h 20 min (20 min per scan). After k-space data reconstruction (slice and in-plane GRAPPA + POCS), all of the coil-combined thick-slice DWI images were reconstructed using conventional gSlider approach \cite{Setsompop2018}, thereby creating high-resolution (860 ${\mu m}$) isotropic data. $B_1ˆ+$ and $T_1$ corrections were applied with the method of \cite{Congyu2019}}. To account for eddy-current distortion and head motion, the high-resolution diffusion-weighted images (64 $\times$ 4 = 256 in total) were processed with the FSL tool $\texttt{EDDY\_CORRECT}$.  Next, the processed four sets of high-resolution DWI were averaged to create a single, SNR-enhanced DWI dataset. Then, spherical harmonics were used to re-sample the DWI data so that the new q-space directions are equally spaced along a spherical spiral that covers the northern hemisphere of $\mathbb{S}^2$ \cite{Michailovich2011a}. This was done to ensure efficient covering of the sphere for different undersampling factors. The ground-truth  $\bm S$ was then defined as the re-sampled DWI dataset normalized by the reference, high-resolution b0 image. 
\paragraph{q-space undersampling scheme:} 
Four different q-space undersampling patterns with several acceleration factors ($2-5 X$) were generated to assess the ability of the proposed algorithm for signal reconstruction. These schemes, which we called  \emph{Scheme 2-5X} respectively, are illustrated in Fig.\ref{fig:schemes}, and ultimately determine the diffusion sampling mask $\bm \Omega_k$  that is used in Eq. \ref{eq:linear_under} for  $k=1,...,N_{RF} =5$ and for every  acceleration factor.

***FIG. 2 APPEARS NEAR HERE***\\

The standard deviation of the noise term $\bm \eta_k$ was defined so as to produce a spatially-averaged SNR of 20 in the low-resolution b0 reference image \cite{Ning2015}. For the proposed gSlider-SR algorithm, the following parameters were used: $\lambda = 0.02, \lambda_{TV}=0.005, \rho_1=\rho_2=0.01$ and $N_{iter} = 8$.  These values were chosen heuristically (after exhaustive search) to provide best results. 

For each under-sampling case (\emph{Scheme 2X} to \emph{Scheme 5X}), a Monte-Carlo (MC) experiment was run. $N_{MC} = 20 $ realizations of the forward-model of Eq.\ref{eq:linear_under} were generated (e.g., $N_{MC} = 20 $  different statistical noise realizations of $\bm \eta_k$).  For comparison purposes, we complement these results with a direct, high-resolution (HR) case with isotropic resolution of 860 $ {\mu m}$, i.e., no gSlider downsampling operator. The SNR  for the HR case in the b0 image was  $n_z = 4.6 $ times lower (SNR $\approx$ 4.5) than the SNR of the thick-slice DWI sets ${\bm Y}_k$.

 \paragraph{Metrics for quantitative validation:} 
Different performance measures were employed to assess the quality of the reconstructed high-resolution data $\bm{ \hat{S}}$ in comparison to the ground-truth $\bm S$. In particular, we were interested in evaluating the performance of gSlider-SR with respect to:

\begin{enumerate}
	\item \emph{Quality of signal reconstruction.} For each voxel $n$,  we calculated the normalized mean-squared error (NMSE) between the estimate $\bm{ \hat{S}}$ and ground truth $\bm S$:
		\begin{equation}
		  \text{NMSE} = \frac{ {\vert \vert \bm {\hat{s}}_n - \bm s_n  \vert \vert}_2^2 }{ {\vert \vert  \bm s_n  \vert \vert }_2^2 }. 
		  \end{equation}
	\item \emph{Accuracy and precision for diffusion-tensor imaging (DTI)}. We fit a DTI model to the reconstructed  high-resolution dMRI data $\bm{\hat S}$ with LLS fitting. For each voxel $n$, we assessed the accuracy and precision in estimating the fractional anisotropy (FA) with respect to the ground-truth FA  (that derived from $\bm S$).  Furthermore, the angular error (in degrees), $\Delta_{\theta}$, between the principal diffusion directions  was also computed:
	\begin{equation}
	\Delta_{\theta} = \frac{180}{\pi} \arccos { ( \bm{\hat{u}} \cdot \bm u ) }	,
	\label{eq:angular_error}
	\end{equation}
	where $\bm{\hat{u}}$  is the main eigenvector of the tensor that is estimated from $\bm{\hat S}$ and $\bm u$ is that of the diffusion tensor estimated from $\bm S$.
	
	\item \emph{Quality of orientation distribution function (ODF) reconstruction.}
		For each voxel $n$ within the white-matter, the ODF $\bm{\hat{S}}$ was estimated using SR fitting, which was then compared to the ODF from the ground-truth data $\bm S$.  The principal diffusion directions and the number of fiber crossings (fiber peaks) were calculated. For a chosen peak in the ground-truth ODF, the angular error (in degrees) between the direction of that peak, $\bm u$,  and the corresponding direction from the reconstructed ODF $\bm{ \hat{u}}$, was calculated as  in Eq.\ref{eq:angular_error}. Next, a single average angular error per voxel $\Delta_{\theta}$ was computed by averaging all errors from each of the ODF peaks in that voxel.  We also calculated the percentage of false peaks $P_d$ in the white matter region.
\end{enumerate}

\subsubsection{Experiments with in-vivo human brain data} %1h
The proposed gSlider-SR framework was also validated with in-vivo human brain data. In this experiment, we tried to assess how well gSlider-SR performs in a real scenario with undersampled gSlider data, in comparison to the fully-sampled case ($N_q = 64$ directions $\times \;5$ RF-encodings = $320$ acquisitions). 
While not a ground-truth per se (due to the presence of spatially varying noise), the fully-sampled, averaged and hence SNR-enhanced DWI set was used as reference, and the high-resolution data reconstructed with gSlider-SR were compared to this set. To obtain the undersampled data, q-space points were removed from the original acquisition in such way as to obtain uniform coverage of the hemisphere. The following parameter settings were used $\lambda = 0.06$, $\lambda_{TV}=10^{-5}$,  $\rho_1=\rho_2=3$ and $N_{iter} = 8$. 

\paragraph{Accounting for subject motion:}
In-vivo real data typically suffer from subject motion and eddy current distortion, as mentioned in the simulation experiment section. In that setting, rigid motion and eddy distortion were corrected by registering  $\bm S$ with affine transformations. Therefore, the synthetically generated thick-slice DWI sets $\bm Y_k$ were free from motion and eddy distortions. In the real scenario, correcting for motion and eddy current distortion directly in the acquired $\bm Y_k$ is not advisable, since the RF-encoding information along slice direction can get mixed-up. An elegant (but computationally involved) solution to this problem was proposed in \cite{Wang2018}, which accounted for motion between RF-encodings, diffusion directions, and slab acquisitions. 

In this work, we use a simpler iterative solution.  We assume that there exist misalignment (affine deformations) between diffusion volumes only, and that relatively small or negligible motion exists between RF-encoding volumes of the same diffusion-weighted direction. Since the LLS problem of Eq.\ref{eq:LLS} is separable along diffusion directions, the reconstruction of each high-resolution diffusion image is then free from motion artifacts.  Nevertheless, the estimate DWI set $\bm S^{(t+1)}$ should be volume-wise registered before solving problem for Eq.\ref{eq:BPD}, as spherical-ridgelets fitting requires the DWI data to be aligned. Coefficients ${\{\bm c_n^{(t+1)}\}}_{n=1}^N$ are then estimated from a registered DWI dataset, $\mathcal{R}\{ \bm S^{(t+1)}\}$, where registration is performed with the FSL tool $\texttt{FLIRT}$ \cite{Jenkinson2002}.  Next, the synthetically generated image defined by $\bm A\bm c_n^{(t+1)}, n=1,...,N$,  (third summand in Eq.\ref{eq:LLS}) is ``unregistered" with the inverse transformation of $\mathcal{R}$, as the solution $\bm S$ in Eq.\ref{eq:LLS} is assumed to be affected by inter-volume motion, and voxel-wise correspondence is required.  After iterating through this process,  head motion and eddy-current distortion can be corrected using $\mathcal{R}\{ \bm S^{(t_{\text{end}})}\}$ where $t_{\text{end}}$ denotes the last iteration of the algorithm.

\section{Results}

\subsection{Simulation experiments } 
Fig. \ref{fig:DWIsimul} shows an axial, coronal, and sagittal slice of a DWI volume from the fully-sampled ground-truth data $\bm S$, as well as DWI volumes reconstructed with Tikhonov regularization and with gSlider-SR methods respectively (undersampling of 2 X). 
From Fig. \ref{fig:DWIsimul} , it is clear that Tikhonov regularization is not enough to reconstruct an accurate, high-resolution dMRI dataset with half of the q-space samples. Nevertheless, gSlider-SR is able to restore a highly detailed, artifact-free diffusion-weighted image, allowing the possibility of decreasing the acquisition time significantly without sacrificing image quality. 
This can be quantitatively confirmed from the NMSE maps of Fig. \ref{fig:NMSEsimul}, where errors of about 2\%  are seen for whole brain (excluding CSF and ventricles) using gSlider-SR, but much higher errors are seen for a simple Tikhonov method.

***FIG. 3 APPEARS NEAR HERE***\\

***FIG. 4 APPEARS NEAR HERE***\\

While the proposed gSlider-SR method shows robust data reconstruction for 2 X acceleration, we also tested the limits of the proposed method for much higher undersampling (and thereby acceleration) factors. In Fig.\ref{fig:metrics} we report quantitative metrics to evaluate the performance of gSlider-SR for various undersampling factors as well as the results with the fully sampled HR case. As expected, as the undersampling ratio increases, the performance of the gSlider-SR degrades. However, it is interesting to note that even for substantially high acceleration factors (4 X), more accurate and precise DTI parameters can be estimated from the high-resolution diffusion-weighted images reconstructed with the gSlider-SR method in comparison to those directly estimated  from an isotropic, high-resolution (860 ${{\mu}m}$) acquisition, see results with label HR in Fig. \ref{fig:metrics}.(b-c). Similar conclusions can be drawn for the angular error in estimating the principal diffusion direction from DTI as well as the directions of the peaks of the ODF,  along with false/missing peaks results (Fig.\ref{fig:metrics}.(d-f)).

  ***FIG. 5 APPEARS NEAR HERE***\\

\subsection{Experiments with in-vivo real data}

An axial, coronal, and sagittal slice from a reconstructed high-resolution DWI volume is shown in Fig.\ref{fig:DWIreal} for the reference set as well as different implementations of conventional gSlider and gSlider-SR. Moreover, the total time to acquire the thick-slice data to reconstruct such DWI sets is also reported.  It can be seen that conventional gSlider reconstruction from one thick-slice DWI set (scan one in this case) suffers from severe noise. The SNR is enhanced if the alloted scan time is doubled to 40 min, and two DWI volumes reconstructed with gSlider are averaged (see gSlider two averages). The SNR-enhancing effect of the spherical ridgelets-based regularization seems evident in this experiment. In the unaccelerated case, gSlider-SR 1X, the reconstructed DWI volume is substantially less noisy than that obtained with conventional gSlider reconstruction, and even with the gSlider two averages case. Interestingly, the reconstructed volume with gSlider-SR 2X seems to present similar visual quality than the non-accelerated case, suggesting that 10 min may be enough to obtain,  an artifact-free, SNR-enhanced, structure-preserving DWI volume that matches well with the reference data. Finally, as expected, averaging the four reconstructed scans with gSlider-SR 2X (40 min) produces the best results in terms of structural preservation and noise reduction (see zoomed-in area).

***FIG. 6 APPEARS NEAR HERE***\\

Quantitative NMSE maps presented in Fig.\ref{fig:NMSEreal} support the claims made above. Evidently,  the reconstruction quality can be further improved if several thick-slice DWI sets are reconstructed with gSlider-SR, and averaged afterward. With a scan time-limit of 20 min, the gSlider-SR 1X method provides substantially better reconstruction than conventional gSlider (note the reduction of NMSE in white and grey matter). Reconstruction errors for different undersampling (and different scan times) are also shown in Fig.\ref{fig:NMSEreal}. We note that, the data quality in the white matter from a 10-minute gSlider-SR method is quite comparable to that of the 80-minute gSlider four averages. Thus, if one were to account for the improved SNR, the proposed gSlider-SR 10-minute scan provides an 8-fold reduction in acquisition time, thereby making the method much more clinically practical. Color-encoded FA maps estimated with the DWI volumes obtained with gSlider-SR (2 X) also present similar visual quality as that obtained with the gSlider four-averages (Fig.\ref{fig:FAreal}).

***FIG. 7 APPEARS NEAR HERE***\\

***FIG. 8 APPEARS NEAR HERE***\\

Finally, we assess the ability of gSlider-SR to recover angular information from the four reconstructed in-vivo gSlider data scans. The angular error for DTI as well as the directions of the ODF peaks were computed, where the reference set was the gSlider four averages method. Results are shown in Tab. \ref{tab:table_new}. 

***TAB. 1 APPEARS NEAR HERE***\\

Although not a direct comparison on the same datasets, angular errors using gSlider-SR are consistently lower than those reported in \cite{Wu2019}, and comparable to those obtained with simulations  (2 X).

\section{Discussion}
In this work, we proposed an accelerated gSlider reconstruction framework (gSlider-SR) which, by means of complementary sampling in the q- and RF-encoded space, robustly reconstructs whole human brain dMRI data at submillimeter isotropic resolution (860 ${\mu}m$)  within a scan time frame that is substantially shorter than that required for conventional gSlider (4-averages). It is important to note that such high resolution data comes at an SNR that is comparable to 4-averages of standard gSlider data, i.e., the proposed method presents an 8-fold acceleration in acquisition time without compromising signal quality.

Using Monte-Carlo simulations, we demonstrated that gSlider-SR is able to accurately reconstruct, structure-preserving, artifact-free, high-resolution DWI datasets. While an acceleration of 2X gives the best performance in terms of normalized mean-square error, angular error as well as DTI-derived measure of FA, the performance is quite stable even for much higher acceleration factors. Comparison of gSlider-SR with conventional gSlider on realistic in-vivo human brain data demonstrated dramatically improved image quality, with significantly reduced scan time of two (gSlider-SR four averages) to eight (gSlider-SR) times shorter than the reference gSlider four averages dataset.

Below, we also discuss the limitations as well as future directions of this work. In-vivo real data that was used in the experiment section required a motion correction scheme, which we smoothly integrated in the gSlider-SR as an iterative registration step with the popular \texttt{FLIRT} algorithm \cite{Jenkinson2002}. While this approach provided very good results, motion and eddy correction can be explicitly modeled within the forward-model of Eq.\ref{eq:linear_under} as is done in \cite{Scherrer2012,Ramos-Llorden2017, Cordero-Grande2018, Wang2018}. This will ensure that the super-resolution DWI dataset and the motion parameters, which vary not only for each diffusion direction but also along with RF-encoding profiles, can be simultaneously estimated within an integrated framework, improving the performance of gSlider-SR \cite{Fogtmann2014,Ramos-Llorden2017}. 

In this work, for simplicity, $\lambda$ was kept constant all over the brain. A spatially varying $\lambda$ could provide a more accurate reconstruction, especially one that can account for spatially varying noise in the image. Indeed, while spherical-ridgelets can model any dMRI signal, the level of sparsity in the gray matter is different than that in white matter. It makes sense then to have a different value of $\lambda$ in gray matter tissue. Interestingly, gSlider-SR can easily include other constrained conditions on the diffusion signal as well as other regularizations terms in the cost function of Eq.\ref{eq:problem}. In particular, we envisage an improved image quality reconstruction due to further noise reduction when spherical ridgelets modeling is combined with low-rank matrix denoising approaches \cite{Manjon2013, Veraart2016,Veraart2016a,Corderogrande2019} and more complex spatial smoothness functionals than simple TV regularization \cite{Haldar2019}.  This research line will be developed in our group in the near future. It should be noted as well that the proposed gSlider-SR reconstruction framework is not confined to single shell but can accommodate multi-shell schemes as well, as spherical-ridgelets have been properly modified for multi-shell diffusion MRI data recovery \cite{Rathi2014}. Finally, the regularization parameters currently have to be chosen manually using a heuristic approach. Our future work entails developing algorithms that do not require manual parameter selection.

\section{Conclusion }

In this work, we have shown that in-vivo diffusion MRI (64 directions with b = 2000 s/$mm^2$) of a whole-brain at isotropic resolution of 860 ${\mu m}$ can be obtained in a clinically feasible scan time with our novel gSlider-SR method. gSlider-SR extends conventional gSlider by allowing both undersampled RF-encoding and q-space data, thereby substantially accelerating the acquisition time of the traditional gSlider protocol. The method allows submillimeter dMRI acquisitions within a clinically feasible scan time, allowing to probe anatomical details not possible with existing methods. 

\section*{Acknowledgments}

We acknowledge funding support from the following National Institute of Health (NIH) grant: R01MH116173 (PIs: Setsompop, Rathi).

%% Bibliography 

%%HERE TABLES, LIST OF FIGURES AND FIGURES ARE SHOWN
\clearpage

\section*{Figures and Table captions}
\noindent 
{\bf Figure 1:} In a conventional gSlider acquisition (a) all the thick-slice DWI sets $\bm Y_k$ probed at the $N_q =64 $ q-space points (dark points) are encoded with the five RF-encoding profiles (vertical axis). However, in the undersampled gSlider acquisition (b), an incomplete RF-encoding basis is used to encode the thick-slices DWI volumes $\bm Y_k$. In this example, DWI volumes that correspond to ``red" q-space points   are encoded only with the first, third, and fifth RF-encoding profile (see vertical axis), whereas DWI volumes probed with ``blue" q-space points  are encoded with the second and fourth RF-encoding profile. This represents an undersampling by a factor of 2. Therefore, the total acquisition time is reduced by half.\\

\noindent 
{\bf Figure 2:} Undersampled q-space schemes that are used in the MC-based simulation experiment. (a)  2 X: DWI volumes probed with red q-space points are encoded with the first, the third and the fifth RF-profile, whereas blue q-space points are encoded with the second and the fourth. (b)  3 X: red, blue and green q-space points are encoded with the first and the fourth, the second and the fifth, and the third RF-encoding profile, respectively. (c) 4 X: red, blue, green, and magenta q-points are encoded with the first and the fifth, the second, the third, and the fourth RF-encoding profile, respectively. (d) 5 X: red, blue, green, magenta and black q-space points are encoded with the first, the second, the third, the fourth, and the fifth RF-encoding profile, respectively.\\

\noindent 
{\bf Figure 3:} Simulation experiment with an acceleration factor of 2 X. A middle axial, coronal and sagittal slice of the same diffusion-weighted volume are shown for the ground-truth $\bm S$ (top row), Tikhonov-based reconstruction (middle row), and gSlider-SR -based reconstruction (bottom row).\\

\noindent 
{\bf Figure 4:} Simulation experiment with an acceleration factor of 2 X. A middle axial, coronal and sagittal slice of the NMSE maps from the reconstructed volumes are shown for the Tikhonov-based reconstruction (top row) and gSlider-SR -based reconstruction (bottom row). \\

\noindent 
{\bf Figure 5:} Quantitative validation of gSlider-SR reconstruction based on a MC-based simulation experiment for different undersampling schemes (2-5 X). Results for the direct, $860 {\mu}m$ isotropic resolution acquisition (HR) are also shown. \\

\noindent 
{\bf Figure 6:} In-vivo data experiment with an acceleration factor of 2 X. A middle axial, coronal and sagittal slice of the same diffusion-weighted volume are shown for gSlider, and gSlider-SR based reconstruction, respectively. \\

\noindent 
{\bf Figure 7:} In-vivo data experiment with an acceleration factor of 2 X. A middle axial, coronal and sagittal slice of the NMSE maps from the reconstructed  DWI volumes are shown for the gSlider, and gSlider-SR based reconstruction, respectively.\\

\noindent 
{\bf Figure 8:} In-vivo data experiment with an acceleration factor of 2 X. A middle axial, coronal and sagittal slice of the color-encoded FA maps estimated from the reconstructed DWI volumes with gSlider (four averages) and gSlider-SR are shown. \\

\noindent
{\bf Table 1:} Quantitative validation of gSlider-SR reconstruction with in-vivo data (acceleration factor of 2 X). The gSlider four averages set was used as reference set.

\clearpage
\section*{Figures} 

\begin{figure}[h]
	\centering
	\includegraphics{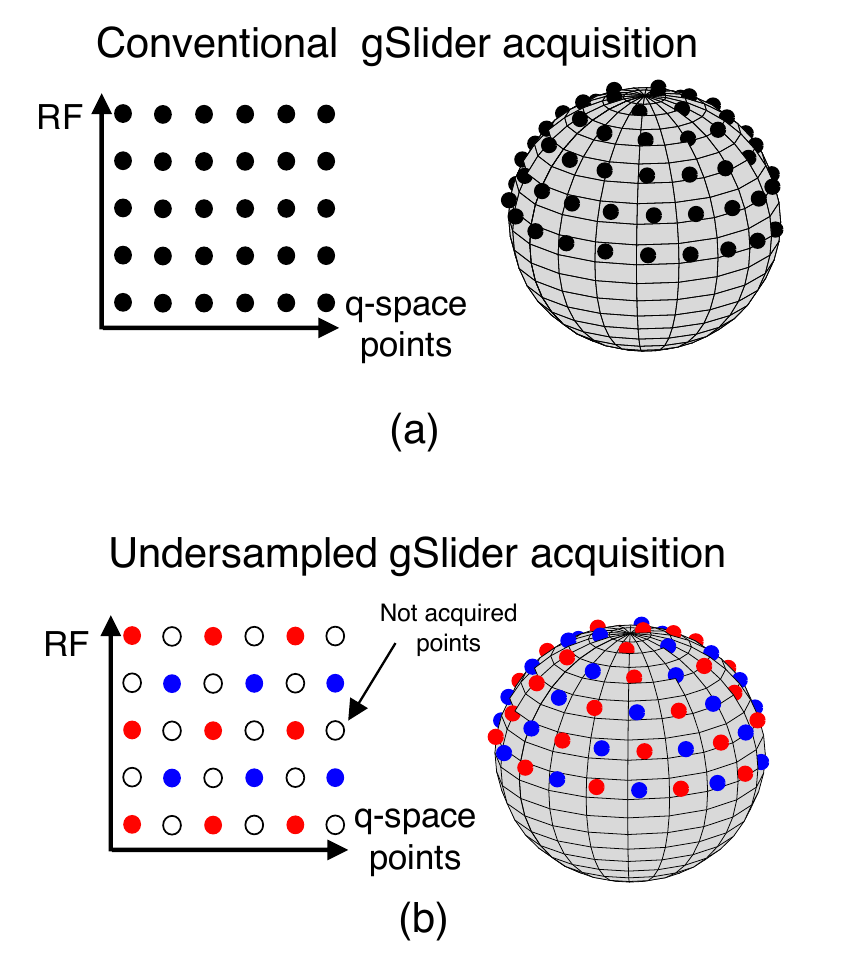} 
	\caption{In a conventional gSlider acquisition (a) all the thick-slice DWI sets $\bm Y_k$ probed at the $N_q =64 $ q-space points (dark points) are encoded with the five RF-encoding profiles (vertical axis). However, in the undersampled gSlider acquisition (b), an incomplete RF-encoding basis is used to encode the thick-slices DWI volumes $\bm Y_k$. In this example, DWI volumes that correspond to ``red" q-space points   are encoded only with the first, third, and fifth RF-encoding profile (see vertical axis), whereas DWI volumes probed with ``blue" q-space points  are encoded with the second and fourth RF-encoding profile. This represents an undersampling by a factor of 2. Therefore, the total acquisition time is reduced by half.}\label{fig:scheme}
\end{figure}

\begin{figure}[h]
	\centering
	\includegraphics{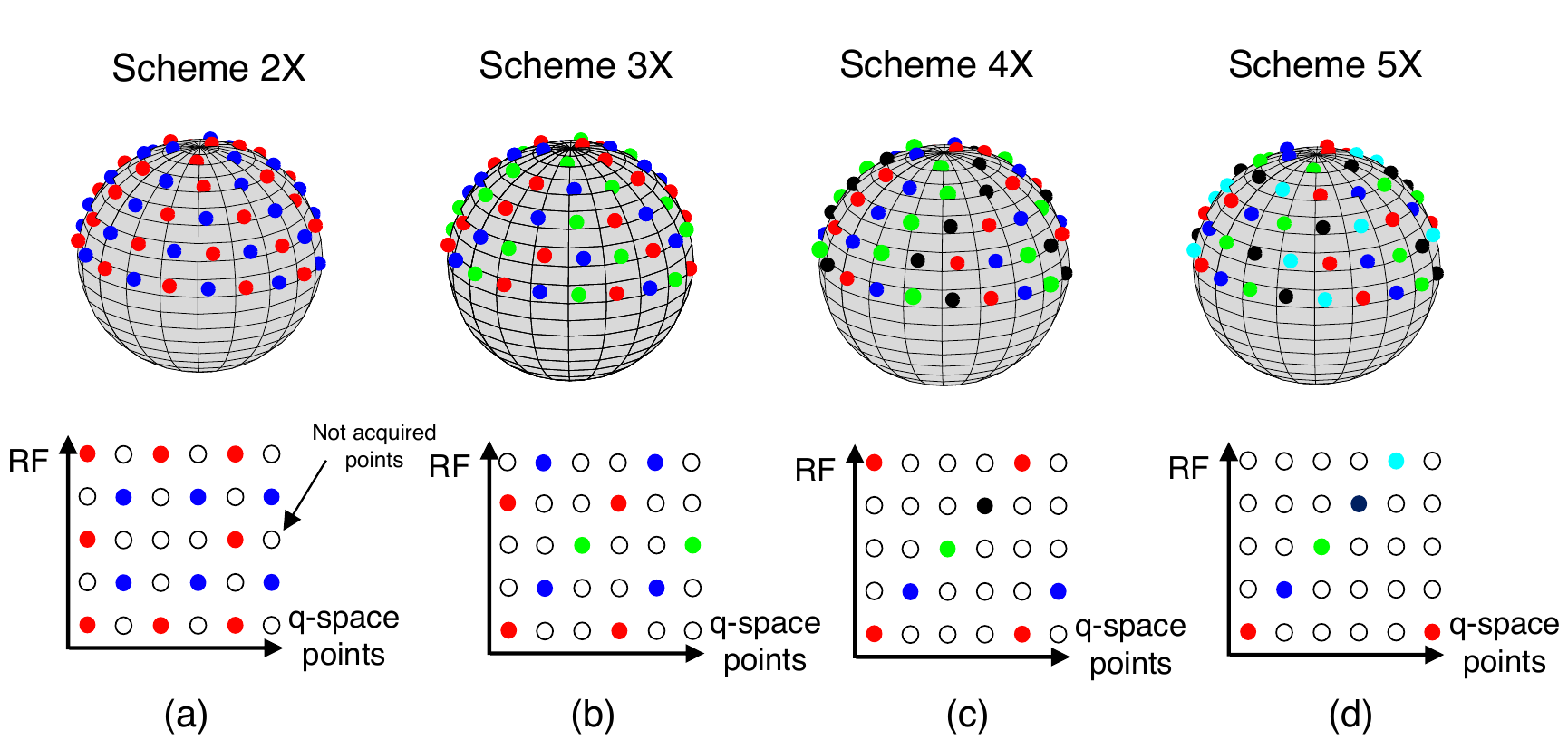} 
	\caption{Undersampled q-space schemes that are used in the MC-based simulation experiment. (a)  2 X: DWI volumes probed with red q-space points are encoded with the first, the third and the fifth RF-profile, whereas blue q-space points are encoded with the second and the fourth. (b)  3 X: red, blue and green q-space points are encoded with the first and the fourth, the second and the fifth, and the third RF-encoding profile, respectively. (c) 4 X: red, blue, green, and magenta q-points are encoded with the first and the fifth, the second, the third, and the fourth RF-encoding profile, respectively. (d) 5 X: red, blue, green, magenta and black q-space points are encoded with the first, the second, the third, the fourth, and the fifth RF-encoding profile, respectively.}\label{fig:schemes}
\end{figure}

\begin{figure}[h]
	\centering
	\includegraphics{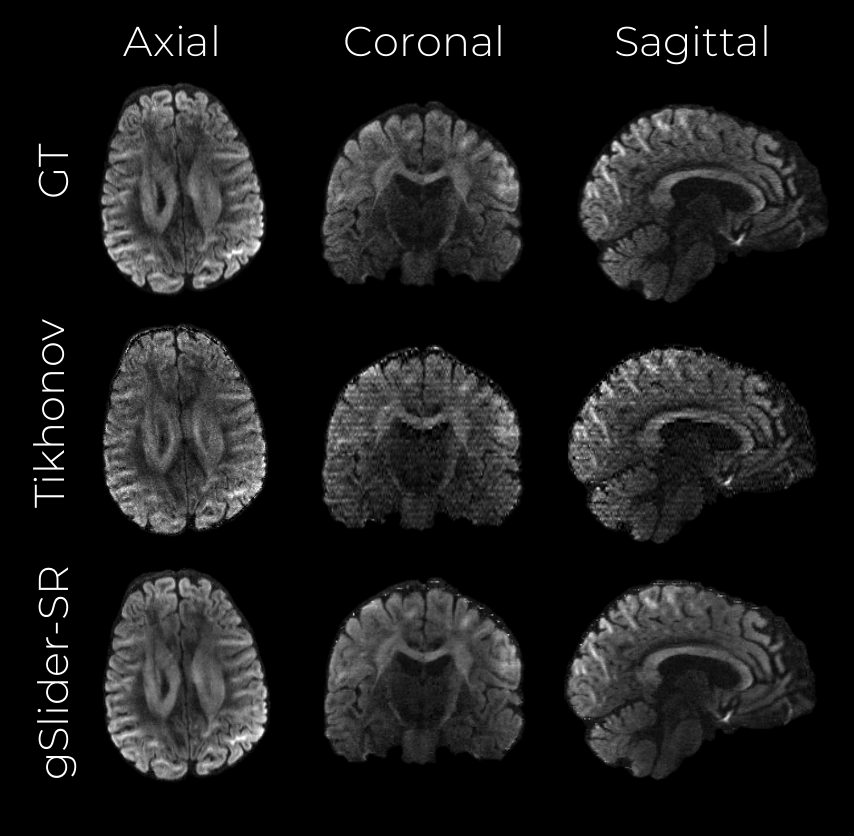} 
	\caption{Simulation experiment with an acceleration factor of 2 X. A middle axial, coronal and sagittal slice of the same diffusion-weighted volume are shown for the ground-truth $\bm S$ (top row), Tikhonov-based reconstruction (middle row), and gSlider-SR -based reconstruction (bottom row).}\label{fig:DWIsimul}
\end{figure}

\begin{figure}[h]
	\centering
	\includegraphics{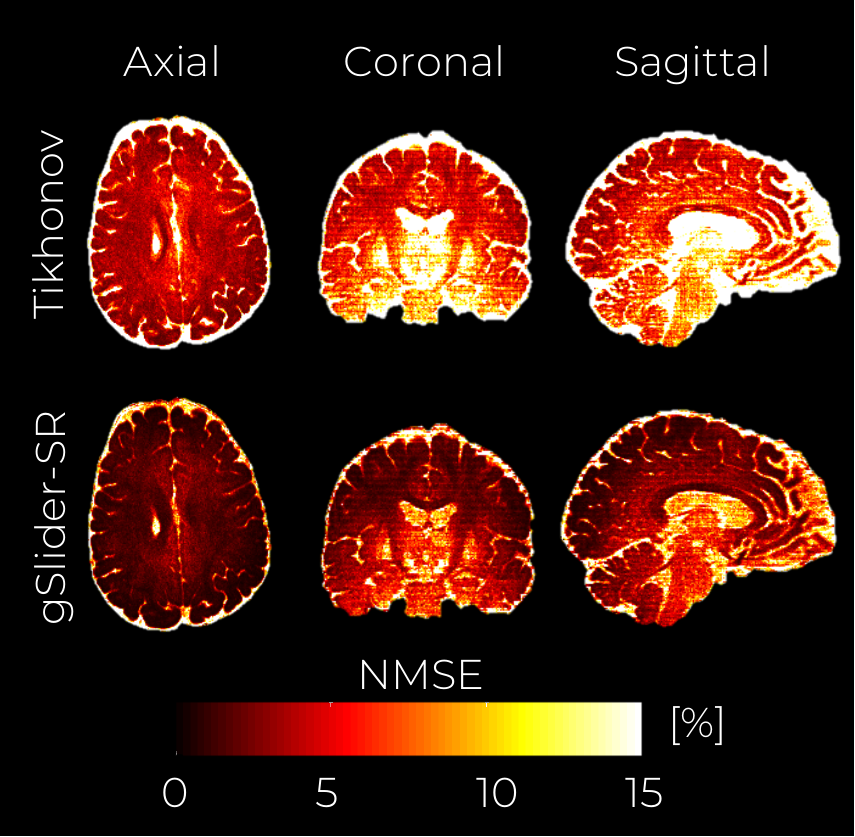} 
	\caption{Simulation experiment with an acceleration factor of 2 X. A middle axial, coronal and sagittal slice of the NMSE maps from the reconstructed volumes are shown for the Tikhonov-based reconstruction (top row) and gSlider-SR -based reconstruction (bottom row).}\label{fig:NMSEsimul}
\end{figure}

\begin{figure}[h]
	\centering
	\includegraphics{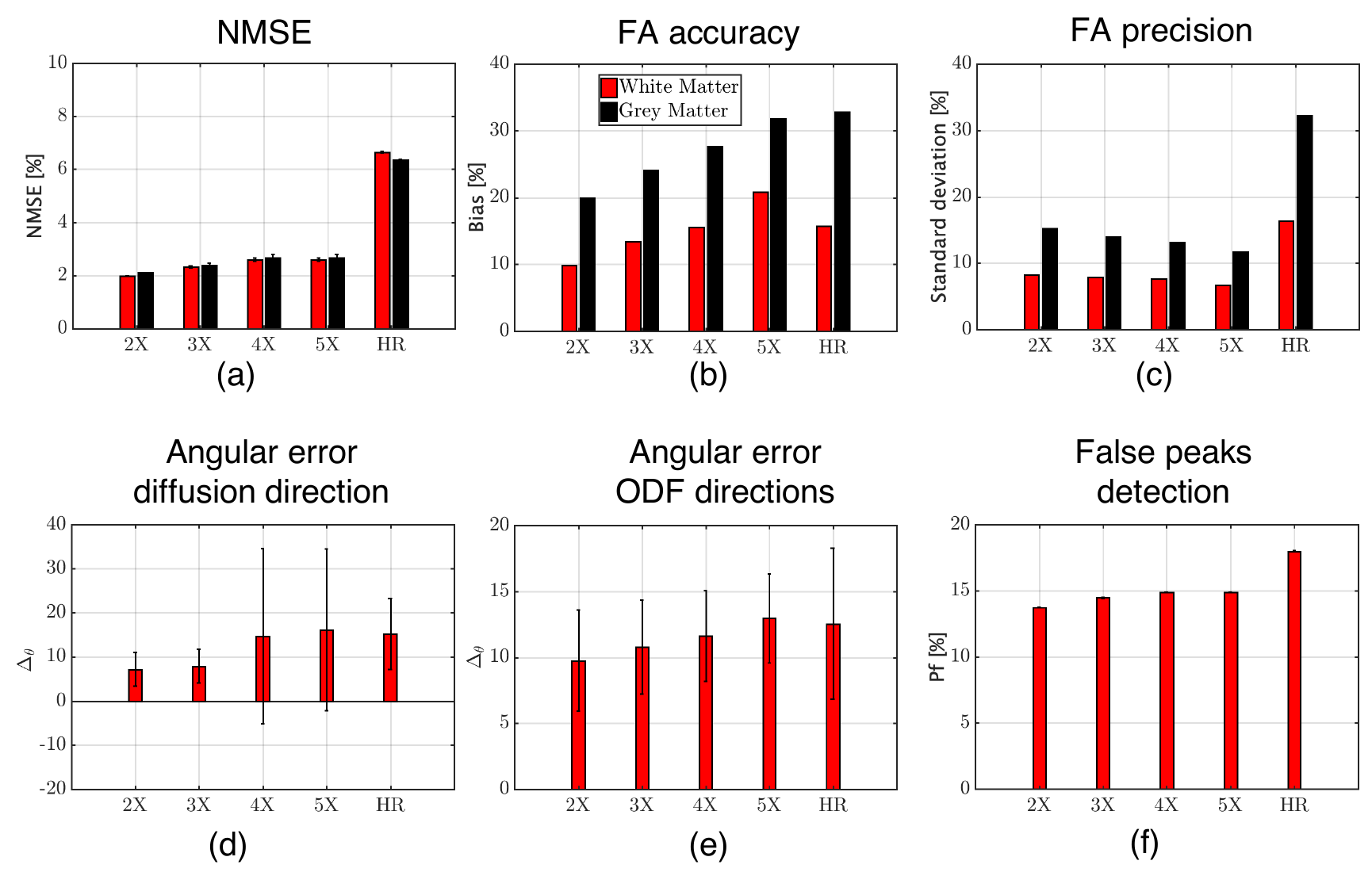} 
	\caption{Quantitative validation of gSlider-SR reconstruction based on a MC-based simulation experiment for different undersampling schemes (2-5 X). Results for the direct, $860 {\mu}m$ isotropic resolution acquisition (HR) are also shown.} \label{fig:metrics}
\end{figure}

\begin{figure}[h]
	\centering
	\includegraphics{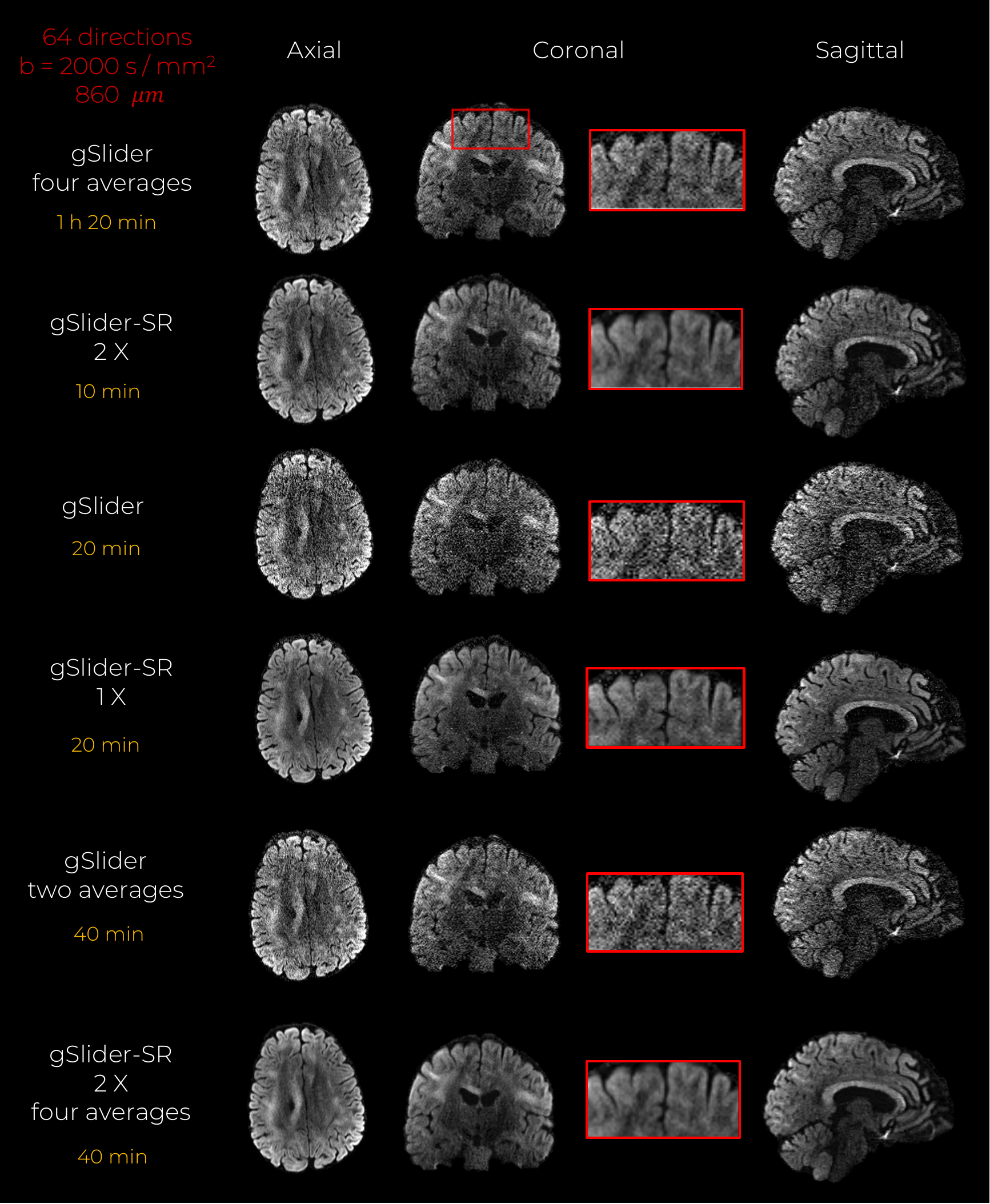} 
	\caption{In-vivo data experiment with an acceleration factor of 2 X. A middle axial, coronal and sagittal slice of the same diffusion-weighted volume are shown for gSlider, and gSlider-SR based reconstruction, respectively.}\label{fig:DWIreal}
\end{figure}

\begin{figure}[htb]
	\centering
	\includegraphics{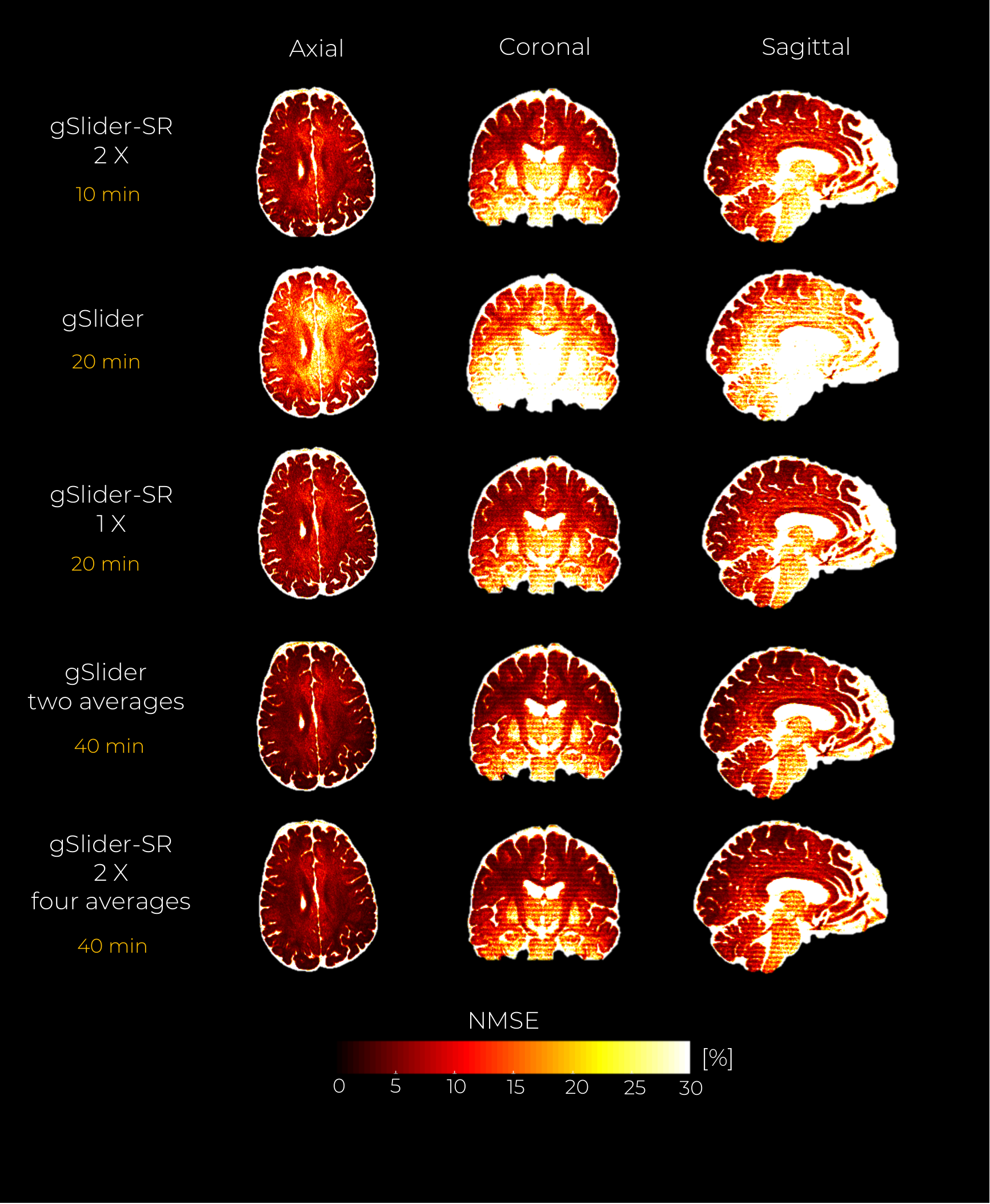} 
	\caption{In-vivo data experiment with an acceleration factor of 2 X. A middle axial, coronal and sagittal slice of the NMSE maps from the reconstructed  DWI volumes are shown for the gSlider, and gSlider-SR based reconstruction, respectively.}\label{fig:NMSEreal}
\end{figure}

\begin{figure}[htb]
	\centering
	\includegraphics{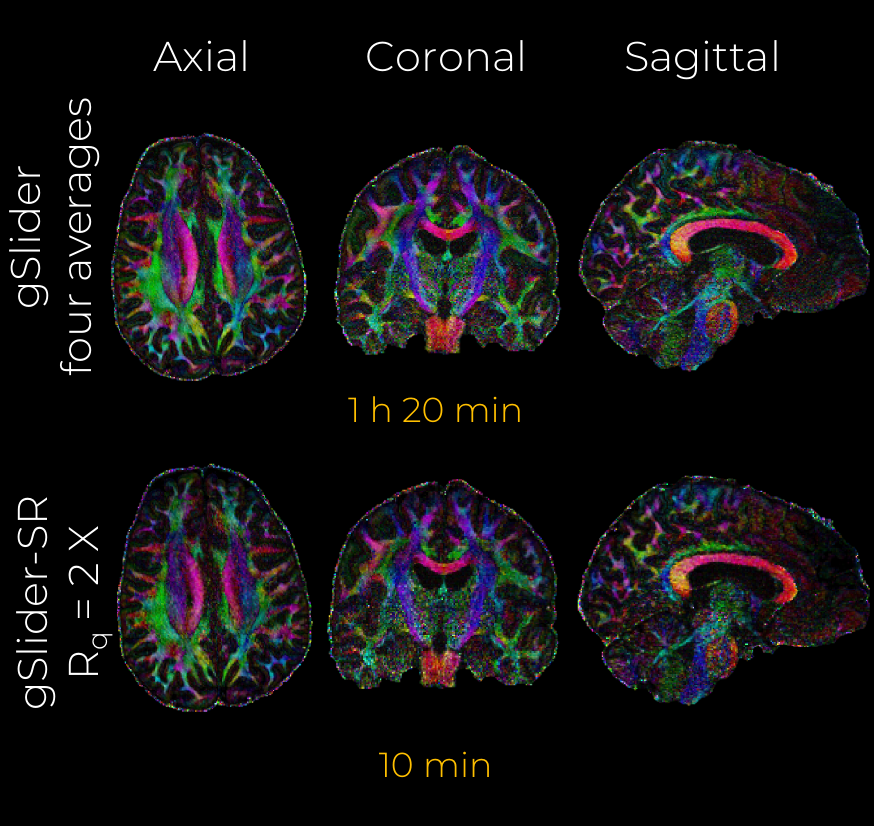} 
	\caption{In-vivo data experiment with an acceleration factor of 2 X. A middle axial, coronal and sagittal slice of the color-encoded FA maps estimated from the reconstructed DWI volumes with gSlider (four averages) and gSlider-SR are shown.}\label{fig:FAreal}
\end{figure}

\clearpage
\section*{Tables} 

\begin{table}[h]
	\centering
	\begin{center}
		{\small
			\begin{tabular}{|l|c|c|c|c|c|c|c|c|c|c|}
				\hline
				\textbf{Metric} & \multicolumn{2}{|c|}{Scan 1} & \multicolumn{2}{|c|}{Scan 2}  &\multicolumn{2}{|c|}{Scan 3}  & \multicolumn{2}{|c|}{Scan 4}   \\ \hline		
				& WM & GM & WM & GM & WM & GM & WM & GM \\ \hline
				NMSE & 10.6 \% & 11.3 \% & 11.2 \% & 12.1 \% & 10.3 \% & 11 \% & 10.4 \% & 11.2 \%  \\ \hline
				& \multicolumn{2}{|c|}{WM} & \multicolumn{2}{|c|}{WM}  &\multicolumn{2}{|c|}{WM}  & \multicolumn{2}{|c|}{WM}   \\ \hline		
				$\Delta_{\theta}$: tensor  & \multicolumn{2}{|c|}{$17.9 ^\circ$} & \multicolumn{2}{|c|}{$18.8 ^\circ$}  &\multicolumn{2}{|c|}{$17.1^\circ$}  & \multicolumn{2}{|c|}{$17.6 ^\circ$}    \\ \hline	
				$\Delta_{\theta}$: ODF peaks   & \multicolumn{2}{|c|}{$16.9 ^\circ$} & \multicolumn{2}{|c|}{$17.3 ^\circ$}  &\multicolumn{2}{|c|}{$16.4 ^\circ$}  & \multicolumn{2}{|c|}{$16.7 ^\circ$}  \\ \hline	
				$P_f$: ODF 	& \multicolumn{2}{|c|}{33 \%} & \multicolumn{2}{|c|}{35.1\%}  &\multicolumn{2}{|c|}{32.4\%}  & \multicolumn{2}{|c|}{32.8\%} \\ \hline			
		\end{tabular}}
	\end{center}
	\caption{Quantitative validation of gSlider-SR reconstruction with in-vivo data (acceleration factor of 2 X). The gSlider four averages set was used as reference set. }\label{tab:table_new}
\end{table}

\newpage

\end{document}